# Sub-5-nm Ultra-thin In$_2$O$_3$ Transistors for High-Performance and Low-Power Electronic Applications

Linqiang Xu,[1,2] Lianqiang Xu,[3] Jun Lan,[4] Yida Li,[4] Qiuhui Li,[1] Aili Wang,[5,6] Ying Guo,[7] Yee Sin Ang,[2*] Ruge Quhe,[8*] and Jing Lu[1,9,10,11,12*]

[1]State Key Laboratory of Mesoscopic Physics and Department of Physics, Peking University, Beijing 100871, China
[2]Science, Mathematics and Technology, Singapore University of Technology and Design (SUTD), 8 Somapah Road, Singapore 487372, Singapore
[3]School of Physics and Electronic Information Engineering, Engineering Research Center of Nanostructure and Functional Materials, Ningxia Normal University, Guyuan 756000, China
[4]School of Microelectronics, Southern University of Science and Technology, 518055, Shenzhen, China
[5]College of Information Science and Electronic Engineering, Zhejiang University, Hangzhou, China
[6]Zhejiang University - University of Illinois at Urbana-Champaign Institute, Zhejiang University, Haining, China
[7]School of Physics and Telecommunication Engineering, Shaanxi Key Laboratory of Catalysis, Shaanxi University of Technology, Hanzhong, 723001, People's Republic of China
[8]State Key Laboratory of Information Photonics and Optical Communications and School of Science, Beijing University of Posts and Telecommunications, Beijing 100876, China
[9]Collaborative Innovation Center of Quantum Matter, Beijing 100871, China
[10]Beijing Key Laboratory for Magnetoelectric Materials and Devices, Beijing 100871, China
[11]Peking University Yangtze Delta Institute of Optoelectronics, Nantong 226000, China
[12]Key Laboratory for the Physics and Chemistry of Nanodevices, Peking University, Beijing 100871, China

*Corresponding Authors: yeesin_ang@sutd.edu.sg, quheruge@bupt.edu.cn, jinglu@pku.edu.cn

## Abstract

Ultra-thin (UT) oxide semiconductors are promising candidates for back-end-of-line (BEOL) compatible transistors and monolithic three-dimensional integration. Experimentally, UT indium oxide (In$_2$O$_3$) field-effect transistors (FETs) with thicknesses down to 0.4 nm exhibits extremely high drain current (10$^4$ μA/μm) and transconductance (4000 μS/μm). Here, we employ the *ab initio* quantum transport simulation to investigate the performance limit of sub-5-nm gate length ($L_g$) UT In$_2$O$_3$ FET. Based on the International Technology Roadmap for Semiconductors (ITRS) criteria for high-performance (HP) devices, the scaling limit of UT In$_2$O$_3$ FETs can reach 2 nm in terms of on-state current, delay time, and power dissipation. The wide bandgap nature of UT In$_2$O$_3$ (3.15 eV) renders it a suitable candidate for ITRS low-power (LP) electronics with $L_g$ down to 3 nm. Both the HP and LP UT In$_2$O$_3$ FETs exhibit superior energy-delay products as compared to other common 2D semiconductors such as monolayer MoS$_2$ and MoTe$_2$. Our study unveils the immense promise of UT In$_2$O$_3$ for both HP and LP device applications.

## 1. Introduction

The downsizing of silicon (Si)-based metal-oxide-semiconductor field-effect transistors (MOSFETs) driven by Moore's law has nearly approached its physical limits due to the short-channel effect (SCE).[1-3] Utilizing channel materials with an ultra-thin body is advantageous for improving gate controllability and mitigating the SCE. Two-dimensional (2D) semiconductors have emerged as promising channel candidates because of their atomic thickness.[4, 5] However, challenges such as high contact resistance, limited large-scale growth technology, and difficulties in forming high-quality dielectrics hinder the further application of 2D semiconductors in transistors.[4, 6-9] Oxide semiconductors can be more advantageous than 2D semiconductors in terms of overcoming the above challenges.[9-12] Oxide semiconductors typically possess a wider bandgap (> 3 eV) than Si (1.12 eV), resulting in reduced leakage current and making them a suitable candidate for low-power (LP) electronics.[13-15] These properties position oxide semiconductors as compelling channel material candidates for next-generation back-end-of-line (BEOL) compatible devices in monolithic 3D integration.

Oxide semiconductors, particularly indium gallium zinc oxide (IGZO), are widely applied in flat-panel display areas.[16] However, their typical characteristics include relatively low carrier mobility ($\mu < 100$ cm$^2 \cdot$V$^{-1} \cdot$s$^{-1}$) and considerable thickness ($t$ of about several tens of nanometers required by the mass production) have limited applications in FET.[17, 18] Fortunately, the recent development of atomic layer deposition method enables the realization of oxide semiconductors, especially the indium oxide (In$_2$O$_3$), with high carrier mobility ($\mu > 100$ cm$^2 \cdot$V$^{-1} \cdot$s$^{-1}$) and sub-1-nm thickness.[10, 19-21] Experimentally, the sub-1-nm-thickness In$_2$O$_3$ FETs exhibit a current *on-off* ratio of 10$^6$, a maximum drain current of several hundred μA/μm, and a maximum transconductance of several hundred μS/μm.[10] Moreover, an extremely high drain current (10$^4$ μA/μm) and transconductance (4000 μS/μm) are achieved in In$_2$O$_3$ FETs with $t$ of 3.5 nm, suggesting the potential of In$_2$O$_3$ in delivering excellent device performance.[22] However, to our best knowledge, the smallest gate length ($L_g$) investigated in ultra-thin In$_2$O$_3$ FETs is 8 nm. The scaling limit of ultra-thin In$_2$O$_3$ FETs, especially in the sub-5 nm $L_g$ regime, has yet to be comprehensively investigated thus far.

In this study, we perform first-principles quantum transport simulation to investigate the transport properties of sub-5 nm UT In$_2$O$_3$ MOSFETs. The underlap (UL) structure offers an effective pathway to improve the device performance by reducing the tunneling current. Using an optimal UL, we show that the on-state current, delay time, and power dissipation of UT In$_2$O$_3$ MOSFETs fulfill the high-performance (HP) demands as outlined in the International Technology Roadmap for Semiconductors (ITRS) for $L_g$ as small as 2 nm. Furthermore, the sizable bandgap of 3.15 eV in UT In$_2$O$_3$ enables its application as LP transistors. Based on the ITRS LP criteria, the scaling limit for UT In$_2$O$_3$ as an LP transistor can reach a gate length of 3 nm. Remarkably, the energy-delay product of UT In$_2$O$_3$ MOSFETs under both HP and LP applications outperforms those of monolayer (ML) MoS$_2$ and MoTe$_2$ MOSFETs. These findings reveal the immense promise of UT In$_2$O$_3$ in both HP and LP electronics.

## 2. Method

We use the QuantumATK 2022 package, which combines non-equilibrium Green's function (NEGF) and density functional theory (DFT), to simulate the device properties of UT In$_2$O$_3$ MOSFETs.[23] The drain current is calculated according to the Landauer–Bűttiker formula:

$$I_{ds} = \frac{2e}{h} \int_{-\infty}^{+\infty} [f_D(E-\mu_D) - f_S(E-\mu_S)] T(E) \, dE \tag{1}$$

here, $\mu_S$ ($\mu_D$) stands for the electrochemical potential of the source (drain), $f_S$ ($f_D$) represents the Fermi-Dirac distribution functions of the source (drain), and $T(E)$ corresponds to the transmission function. The average of the transmission coefficient $T_{k_{//}}(E)$ over the surface-parallel reciprocal lattice vector $k_{//}$ in the irreducible Brillouin zone results in $T(E)$. $T_{k_{//}}(E)$ can be obtained by:

$$T_{k_{//}}(E) = Tr[\Gamma_{k_{//}}^l(E) G_{k_{//}}(E) \Gamma_{k_{//}}^r(E) G_{k_{//}}^\dagger(E)] \tag{2}$$

where $G_{k_{//}}(E)/G_{k_{//}}^\dagger(E)$ is retarded/advanced central Green's function, and $\Gamma_{k_{//}}^{l/r}(E)$ is the broadening matrix. Based on the formula $\Gamma_{k_{//}}^{l/r}(E) = i[\Sigma_{k_{//}}^{l/r} - (\Sigma_{k_{//}}^{l/r})^\dagger]$, $\Gamma_{k_{//}}^{l/r}(E)$ is equal to the imaginary part of self-energy $\Sigma_{k_{//}}^{l/r}$, which reflects the interaction between the channel and the left/right electrodes. The PseudoDojo pseudopotential, density mesh cutoff of 125 Hartree, temperature of 300 K, and $k$-point mesh of 3×1×92 are applied. For the vertical, transverse, and transport directions, we adopt the Neumann, Periodic, and Dirichlet boundary conditions, respectively.

Throughout the simulation, the Perdew-Burke-Ernzerhof (PBE) of the generalized gradient approximation (GGA) is employed.[24] Since the electron-electron interaction is heavily screened by the dielectric layer and injecting carriers from electrodes, the DFT-GGA method is accurate enough to assess the bandgap in the device.[25-27] For example, the dielectric screening results in a bandgap renormalization and agreement in the $HfO_2$ sandwiched ML $MoS_2$ system (bandgap of GAA/GW is 1.76/1.9 eV),[28,29] while the injecting carriers screening leads to a bandgap consistency in the degenerately doped ML $MoSe_2$ system (bandgap of GAA/GW/experiment is 1.52/1.59/1.58 eV)[30-32]. Moreover, the agreement in transfer curves, on-state current, delay time, and power dissipation between experimental and theoretical carbon nanotube MOSFETs with $L_g$ of 5 nm confirms the feasibility of the DFT-NEGF method.[33]

## 3. Results
### 3.1 Device Structure and On-state Current

Cubic bulk $In_2O_3$ is utilized to construct the UT $In_2O_3$, as depicted in Figure 1(a).[22] The thickness of UT $In_2O_3$ is 0.43 nm, and the outermost oxygen layers are passivated by the hydrogen atoms. Figure 1(b) shows the calculated band structure of the UT $In_2O_3$, which has a direct wide bandgap ($E_g$) of 3.0 eV. Notably, this wide bandgap is advantageous for minimizing the leakage current ($I_{leakage} \propto -E_g$),[34] making UT $In_2O_3$ a suitable candidate for LP electronics, as demonstrated below. The electron and hole effective masses ($m^*$) of the UT $In_2O_3$ are extracted from the band structure with the values of 0.436 and 15.85 $m_0$, respectively ($m_0$ is the mass of an electron). The device structure of UT $In_2O_3$ MOSFET is illustrated in Figure 1(c). The source and drain electrodes adopt the heavily doped UT $In_2O_3$, while the channel is simulated by a pristine UT $In_2O_3$. An underlap (UL) structure is added between the gate and each electrode.

There is no standard for the sub-5-nm $L_g$ because the smallest $L_g$ scaling of the ITRS 2013 version and the latest International Roadmap for Devices and Systems (IRDS) 2022 version are 5 and 12 nm, respectively. To address this, we extrapolate the ITRS 2013 criteria for various parameters into the sub-5 nm $L_g$ regime based on the available data, as presented in Tables S1 (HP) and S2 (LP). Hereafter, we refer "ITRS 2013 version" as "ITRS" for notational simplicity. Considering the experimentally observed *n*-type characteristics of the UT $In_2O_3$,[10, 19, 20, 22] electron doping is implemented for the electrodes. After testing the doping concentration (Figure S1), an optimal concentration of $1\times10^{14}$ cm$^{-2}$ is selected for the subsequent simulation. The $I$-$V_g$ curves for HP and LP applications with $L_g$ ranging from 1 to 4 nm are depicted in Figures S2 and S3, respectively.

On-state current ($I_{on}$) is a crucial figure of merit for transistors. The on-state point ($I_{on}$, $V_g^{on}$) at $I$-$V_g$ curve is determined by $V_g^{on} = V_g^{off} + V_{dd}$, where $V_g^{off}$ is the off-state voltage corresponding to $I_{off}$, and $V_{dd}$ is the supply voltage. The ITRS recommended values of $I_{off}$ and $V_{dd}$ are presented in Tables S1 and S2. Figure 2 shows the relationship between $I_{on}$ and $L_g$. For HP devices [Figure 2(a)], $I_{on}$ of the UT $In_2O_3$ MOSFETs exceeds the ITRS demands only at $L_g$ = 4 nm without UL. In contrast, the inclusion of the UL structure significantly enhances the device performance, and further reduces the scaling limit of UT $In_2O_3$ MOSFETs to 2 nm. As to the LP devices (Figure 2(b)), $I_{on}$ is insufficient to meet the ITRS goal without the help of UL structure. By incorporating the UL structure, $I_{on}$ of 3-and 4-nm-$L_g$ UT $In_2O_3$ MOSFETs are significantly increased by $3.3\times10^4$ and 80 times, thus reaching the LP ITRS requirements.

## 3.2 Mechanism of Underlap Structure

As discussed above, the UL structure effectively enhances the $I_{on}$ of UT $In_2O_3$ MOSFETs. To understand the working mechanism of UL structure more clearly, we plot the local density of states (LDOS) and the spectrum current density of UT $In_2O_3$ MOSFETs at $L_g$ = 2 nm (Figure 3). The electron barrier height $\Phi$ that the electron needs to overcome can be extracted from the LDOS. The value of $\Phi$ is defined as the difference between the Fermi level of drain ($\mu_d$) and the conduction band minimum (CBM) of central channel. $\Phi$ is tuned by applying $V_{dd}$ (0.57 V for $L_g$ = 2 nm). The spectrum current density, which contains thermionic current ($I_{therm}$) and tunneling current ($I_{tunnel}$), is also shown in Figure 3 as a function of carrier energy.[35, 36] Here, the $I_{therm}$ and $I_{tunnel}$ are defined as current contributed by electronic states lying above and below the CBM of central channel, respectively.

At the off-state, $I_{tunnel}$ dominates over $I_{therm}$ for both UL lengths of 0 and 1 nm. A longer UL length corresponds to a larger barrier width $w$, leading to a smaller $I_{tunnel}$ ($I_{tunnel} \propto e^{-w\sqrt{m^*\Phi}}$). To achieve the same $I_{off}$ at $L_g$ = 2 nm (0.1 μA/μm), 1-nm UL generally requires a smaller $\Phi$. This is evident in the LDOS, where $\Phi$ for UL = 1 nm (0.23 eV) is smaller than that of UL = 0 nm (0.69 eV). When $V_{dd}$ is applied, $\Phi$ is reduced, and the corresponding UT $In_2O_3$ MOSFETs are switched from off-state to on-state. As the modulation of $\Phi$ is similar for both devices, a considerable $\Phi$ (0.33 eV) is observed for the 0-nm-UL MOSFET at on-state due to the large $\Phi$ at off-state, while $\Phi$ of the 1-nm-UL MOSFET is decreased to -0.15 eV. Therefore, the current density at the on-state is dominated by $I_{tunnel}$ and $I_{therm}$ for UL of 0 and 1 nm, respectively, which leads to a difference of about three magnitudes of order in the peak current density ($10^{-8}$ *vs* $10^{-5}$ A/eV for UL = 0 *vs* 1 nm) and $I_{on}$ (4 *vs* 1020 μA/μm for UL = 0 *vs* 1 nm). In a word, a longer UL length usually results in a smaller $\Phi$ for the off-state and on-state and thus a larger on-state current.

### 3.3 Gate Controllability

A proper UL length is also beneficial for promoting gate control. The UL serves as a spacer that reduces the influence of electrodes on the channel, as observed in the LDOS. When the UT $In_2O_3$ MOSFETs are tuned from the off-state to the on-state, $\Phi$ is decreased by 0.38 eV for UL = 1 nm, which is larger than that of the 0-nm UL (0.36 eV). The gate controllability in the subthreshold region can be described by the subthreshold swing $SS = \frac{\partial V_g}{\partial \lg I}$, as shown in Figures 4(a) and 4(b) for HP and LP devices, respectively. A smaller $SS$ indicates superior gate control. As expected, the addition of the UL structure leads to a reduction in $SS$ for both the HP and LP devices. The decreasing ratio of the HP UT $In_2O_3$ MOSFET is (33, 69) % at $L_g$ = (4, 3) nm, respectively, while it turned out to be (22, 43, 68) % at $L_g$ = (4, 3, 2) nm for the LP UT $In_2O_3$ MOSFET, respectively. The smallest $SS$ of HP and LP can reach 77 and 66 mV/dec, respectively, which approaches the room-temperature SS limit of 60 mV/dec.

The gate controllability promotion is also found in the superthreshold region, which can be illustrated by the transconductance $g_m = \frac{dI}{dV_g}$. A larger $g_m$ is preferred. Figures 4(c) and 4(d) depict $g_m$ of the HP and LP UT $In_2O_3$ MOSFETs as a function of $L_g$. For HP devices, the assistance of UL structure increases $g_m$ by (5.7, 8.5)% for $L_g$ = (4, 3) nm and 230 times for $L_g$ = 2 nm. In the case of LP devices, $g_m$ of the 4-and 3-nm UT $In_2O_3$ MOSFETs are smaller than 100 μS/μm. In contrast, by adding the UL, $g_m$ is enhanced to 3264 and 2865 μS/μm, respectively. These findings further demonstrate the improvement in gate controllability. Additionally, the largest $g_m$ values reach 3415 and 3264 μS/μm for HP and LP devices, respectively. Therefore, both the gate control abilities in the subthreshold and superthreshold regions are significantly improved by the UL structure.

### 3.4 Delay Time, Power Dissipation, and Energy-Delay Product

The delay time $\tau$ depicts the switching speed of a FET and can be obtained via $\tau = C_t V_{dd}/I_{on}$, where $C_t$ is the total capacitance. A smaller $\tau$ is preferred. According to the ITRS criteria, $C_t$ is three times $C_g$, which is defined as $C_g = \partial Q_{ch}/\partial V_g$ ($Q_{ch}$ is the total charge in the channel). The $C_t$ of UT $In_2O_3$ MOSFETs with respect to $L_g$ is plotted in Figure S4. For both HP and LP devices, the optimal $C_t$ reaches the ITRS target with $L_g$ down to 1 nm. Owing to the improvement in $I_{on}$ by the UL structure, the $\tau$ of UT $In_2O_3$ MOSFETs is observed to decrease with UL [see Figures 5(a) and 5(b)]. The UL-optimized $\tau$ ranges from 0.095 to 0.141 ps ($L_g$ at 1-4 nm) and 0.256 to 0.505 ps ($L_g$ at 3-4 nm) for HP and LP devices, respectively, fulling the ITRS criteria.

Another significant indicator in assessing power cost is the power dissipation PDP = $V_{dd} I_{on} \tau = C_t V_{dd}^2$. A smaller PDP indicates lower power consumption. The PDPs of HP and LP UT $In_2O_3$ MOSFETs are shown in Figures 5(c) and 5(d), respectively. Since PDP is proportional to $C_t$, a similar trend of $C_t$ and $L_g$ is observed. The scaling limits of PDP for HP and LP applications can reach 1 nm according to the ITRS criteria. Considering $\tau$ and PDP simultaneously, the energy-delay product EDP = $\tau$ × PDP represents another important figure of merit for MOSFET. Figure 6 compares EDP of the UT $In_2O_3$ MOSFETs with that of other 2D materials (ML $MoS_2$, $MoTe_2$, GeSe, and $ReS_2$) MOSFETs at $L_g$ of 1-5 nm.[37-40] For the HP devices, UT $In_2O_3$ exhibits superior EDP than ML $MoS_2$ and $MoTe_2$ at the same $L_g$, while it possesses comparable EDP with ML GeSe along both armchair and zigzag directions. As to the LP devices, EDP of the UT $In_2O_3$ MOSFETs exceeds almost all the other 2D material counterparts. These results highlight the excellent performance of UT $In_2O_3$ MOSFETs in terms of EDP.

## 4. Discussion

We now discuss the potential experimental realization of UT In$_2$O$_3$ MOSFET in terms of device fabrication process. A possible fabrication process flow is illustrated in Figure 7. The device can be fabricated on a Si substrate with a 285 nm-SiO$_2$ thermally grown onto the surface. The bottom gate consisting of Ti/Pt (5/20nm) is then deposited by e-beam evaporation (EBE). Based on the atomic layer deposition (ALD) method, the bottom dielectric layer (UT HfO$_2$) and the channel layer (UT In$_2$O$_3$) can be subsequently deposited. The thicknesses can be confirmed by ellipsometer. Considering the difficulty of forming an ultra-short channel length by one-step e-beam lithography, the deposition of source/drain electrodes is divided into two steps. The pattern of source electrode is defined first and that of drain electrode is defined next. EBE is then used to deposit the metals for both steps. To form a dual gate structure, the UT HfO$_2$ is deposited again as the top dielectric layer using ALD, while the top gate (Ti/Pt, 5/20nm) can then be defined by lithography and deposited using EBE, yielding a UT In$_2$O$_3$ MOSFET.

## 5. Conclusion

In conclusion, we simulated the transport characteristics of a sub-5 nm UT In$_2$O$_3$ MOSFET by combing DFT and NEGF methods. A proper UL length is useful for improving the device performance. By employing an appropriate UL, the $I_{on}$, $\tau$, and PDP of the UT In$_2$O$_3$ MOSFETs can achieve the ITRS HP targets with $L_g$ scaled down to 2 nm. Additionally, the 3- and 4-nm $L_g$ UT In$_2$O$_3$ MOSFETs can meet the ITRS requirements for LP devices – a direct consequence of the wide bandgap of In$_2$O$_3$. UT In$_2$O$_3$ outperforms the ML MoS$_2$ and MoTe$_2$ counterparts in terms of EDP. These findings highlight the significant potential of UT In$_2$O$_3$ in both the HP and LP applications, and may pave a way towards the next-generation of sub-5-nm device technology.

**Acknowledgment**

This work is supported by the National Natural Science Foundation of China (No. 91964101, No. 12274002, and No. 12164036, No. 62174074), the Ministry of Science and Technology of China (No. 2022YFA1200072, No.2022YFA1203904), the China Scholarship Council, the Fundamental Research Funds for the Central Universities, the Natural Science Foundation of Ningxia of China (No. 2020AAC03271), the youth talent training project of Ningxia of China (2016), and the High-performance Computing Platform of Peking University. Y. L and J. H acknowledge the supports of National Natural Science Foundation of China (Grant No. 62174074), Shenzhen Fundamental Research Program (Grant No. JCYJ20220530115014032), Zhujiang Young Talent Program (Grant No. 2021QN02X362), Guangdong Provincial Department of Education Innovation Team Program (2021KCXTD012). Y. S. A. and Linqiang Xu acknowledge the supports of Singapore University of Technology and Design Kickstarter Initiatives (SKI) under the Award No. SKI 2021_01_12 and SUTD Startup Research Grant (SRG SCI 2021 163). Y. S. A. and A. W. are also supported by the SUTD-ZJU IDEA Thematic Research Grant Exploratory Project (SUTD-ZJU (TR) 202203). The computational work for this article was partially performed on resources of the National Supercomputing Centre, Singapore (https://www.nscc.sg).

**Conflict of Interest**

The authors declare no conflict of interest.

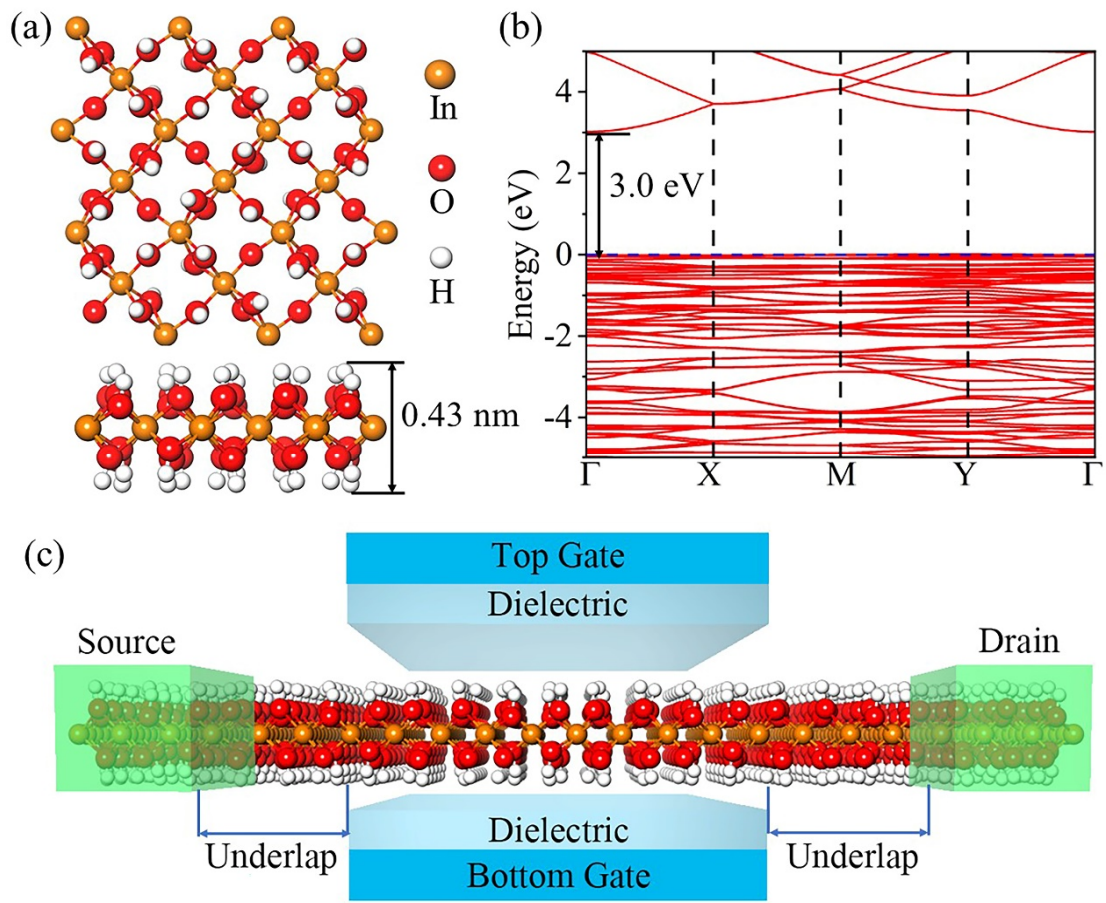

**Figure 1.** (a) Top and Side views of the UT $In_2O_3$. The thickness of UT $In_2O_3$ is 0.43 nm. (b) Band structure of the UT $In_2O_3$. The Fermi level is set at the valence band maximum (0 eV) and represented by the blue dashed line. (c) Schematic diagram of the UT $In_2O_3$ MOSFET.
10

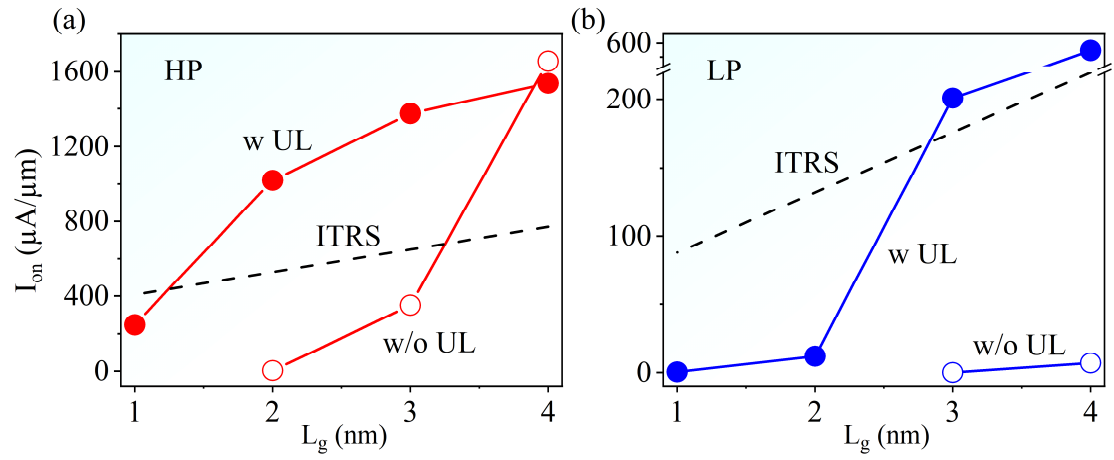

**Figure 2.** $I_{on}$ of the UT $In_2O_3$ MOSFETs versus $L_g$ for (a) HP and (b) LP applications. The solid and hollow circles indicate with (w) and without (w/o) UL structures, respectively. $I_{on}$ with UL structure is the optimal value. The dashed line stands for the ITRS standard.

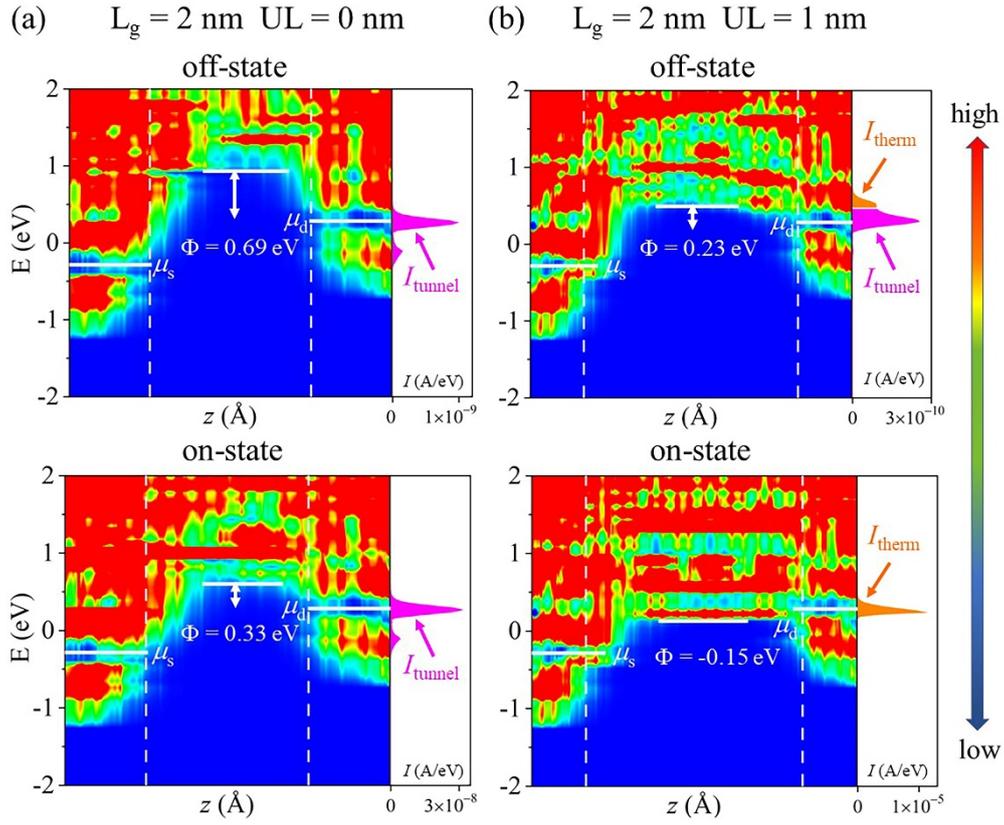

**Figure 3.** Local density of states (LDOS) and spectrum current of the UT $In_2O_3$ MOSFETs at $L_g$ = 2 nm and (a) UL = 0 nm, (b) UL = 1 nm. $\Phi$ represents the electron barrier height, and $\mu_s$ ($\mu_d$) stands for the Fermi level of the source (drain). The white dashed line indicates the boundary between the electrode and channel regions.

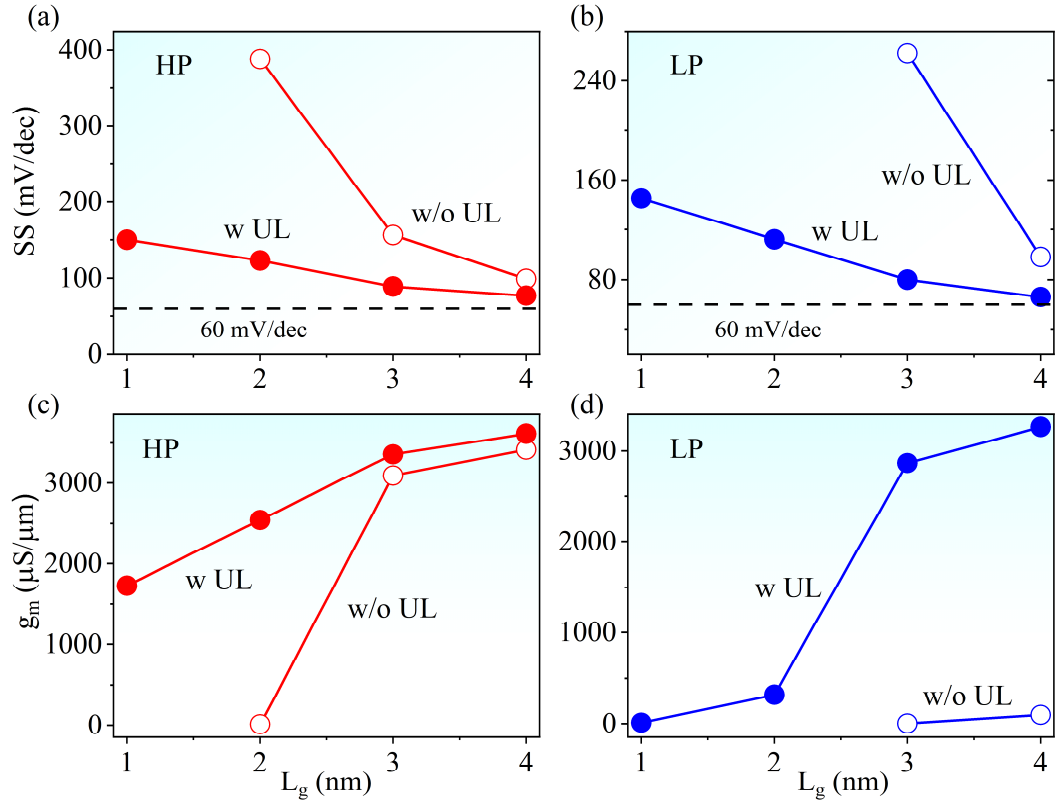

**Figure 4.** Subthreshold swing and transconductance of the UT $In_2O_3$ MOSFETs as a function of $L_g$ for (a) HP *SS*, (b) LP *SS*, (c) HP $g_m$, and (d) LP $g_m$. The solid and hollow circles represent with (w) and without (w/o) UL structures, respectively. *SS* and $g_m$ with UL structure are the optimal values. The black dashed lines in (a) and (b) indicate the *SS* limit at room temperature.

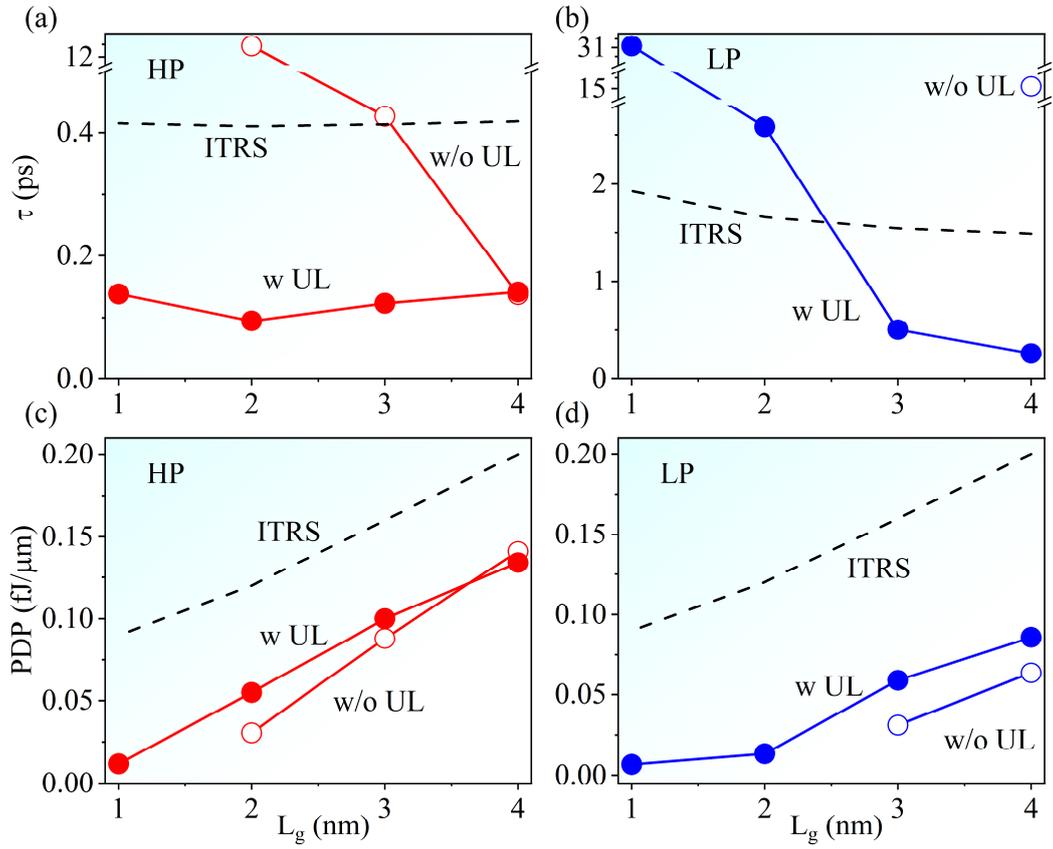

**Figure 5.** Delay time and power dissipation of the UT $In_2O_3$ MOSFETs as a function of $L_g$ for (a) HP $\tau$, (b) LP $\tau$, (c) HP PDP, and (d) LP PDP. The solid and hollow circles represent with (w) and without (w/o) UL structures, respectively. $\tau$ and PDP with UL structure are the optimal values. The dashed line stands for the ITRS standard.

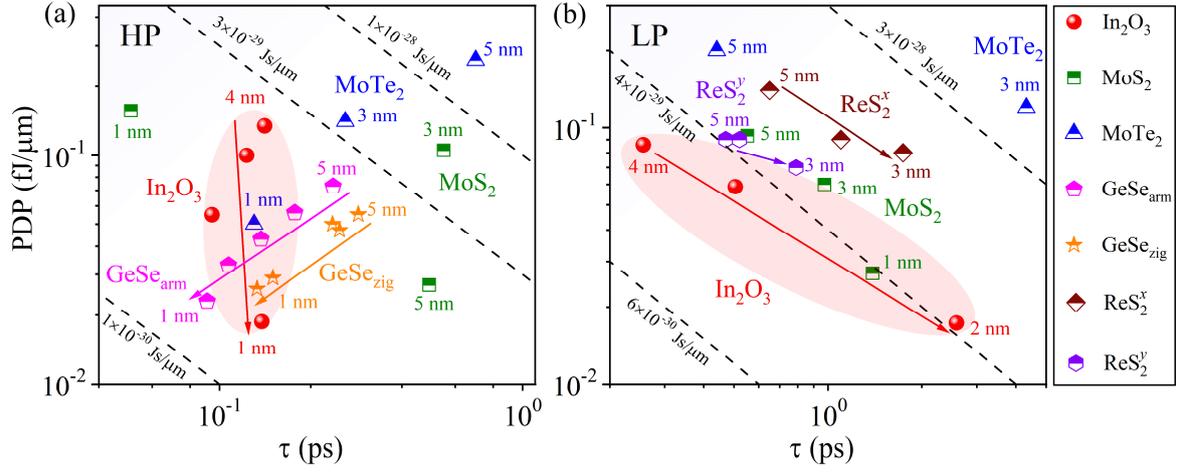

**Figure 6.** UL-optimized PDP versus $\tau$ of the (a) HP and (b) LP devices for UT $In_2O_3$, ML $MoS_2$, ML $MoTe_2$, ML GeSe along armchair direction ($GeSe_{arm}$), ML GeSe along zigzag direction ($GeSe_{zig}$), ML $ReS_2$ along $x$ direction ($ReS_2^x$), and ML $ReS_2$ along $y$ direction ($ReS_2^y$) MOSFETs at $L_g$ = 1-5 nm.[37-40] The dashed lines indicate different values of the energy-delay product EDP = $\tau \times$ PDP.

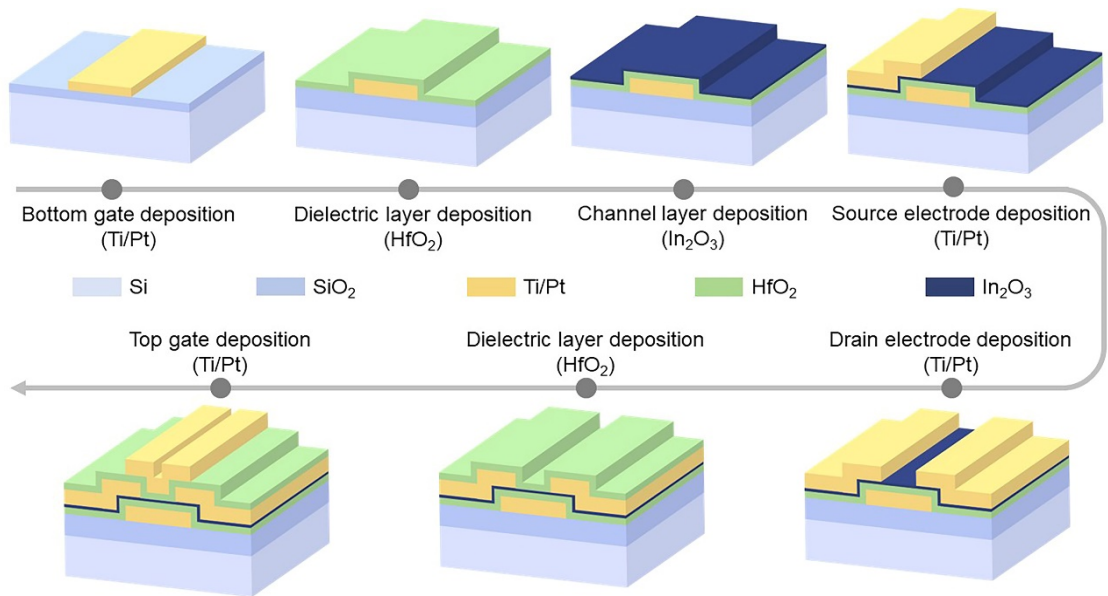

**Figure 7.** Proposed fabrication process flow of the UT $In_2O_3$ MOSFETs.